\documentclass[aps,prl,twocolumn,amsmath,amssymb,superscriptaddress]{revtex4-2} 
\usepackage{amsmath}
\usepackage{graphicx} 
\usepackage{dcolumn} 
\usepackage{bm} 
\usepackage{tikz} 
\usepackage{pgfplots} 
\pgfplotsset{compat=1.15} 
\usetikzlibrary{decorations.pathmorphing,decorations.markings,arrows.meta} 
\usepackage[colorlinks=true,citecolor=blue,linkcolor=red,urlcolor=blue]{hyperref}

\begin{document}

\title{Gravitationally Induced Entanglement Across an Event Horizon}

\author{Hatim Salih} 
\email{salih.hatim@gmail.com}
\affiliation{York Centre for Quantum Technologies, University of York, Heslington, York YO10 5DD, United Kingdom}

\date{\today}

\begin{abstract}
It has long been assumed that a particle---having crossed a black hole event horizon unentangled with another particle in the exterior universe---can no longer dynamically entangle with that particle. Here, we first establish the possibility of post-crossing entanglement via an electromagnetic mediator. We then show within a gravitational time-delayed-potential model that entanglement can be created between freely falling spatial superpositions across an event horizon. Radial escape, however, triggers bremsstrahlung emission---imposing a dephasing bound that grows with entangling phase, $\Gamma > \frac{6}{11\pi}\Phi$, within a constrained leading-quadrupole model---and is ultimately ruled out by macroscopic which-path leakage. By contrast, equivalent macroscopic optical masses in principle allow local harvesting of entanglement tangentially, via quantum erasure, thus revealing a remarkable geometric duality: Spacetime irreversibly degrades entanglement under radial escape, while allowing the transverse teleportation of the entangled qubit to infinity, independent of Hawking evaporation.
\end{abstract}

\maketitle

\textit{Introduction.}---A black hole event horizon acts as an absolute one-way causal boundary \cite{Wald1994}: no physical influence can propagate outward from the black hole interior. Consequently, if a particle crosses the horizon unentangled with an exterior probe, the local bipartite interaction required for dynamically generating entanglement seems permanently lost. While one can distribute pre-existing entanglement across an event horizon (such as the in-falling half of an entangled pair), that would merely constitute kinematic transport. By contrast, the dynamical generation of bipartite entanglement across a null boundary is assumed to be causally forbidden.

Separability of the probes, however, does not imply separability from the mediating field. Correlations established before crossing can subsequently be converted, as we show, into probe--probe entanglement, even when the probes no longer exchange signals in both directions.

For freely falling parties Alice and Bob near a Schwarzschild horizon, the equivalence principle of Einstein ensures that the local spacetime remains flat, provided the apparatus and protocol duration remain small compared to local curvature scales \cite{Wald1994,Einstein1907}. Each party holds a confined neutral probe in a superposition of two opposite tangential electric-dipole configurations, unitarily mapped to internal states $|L\rangle,|R\rangle$ \cite{ChargeDeMille2002,ChargeYelin2006}. Their quantum electromagnetic coupling continues after Bob crosses. The pre-crossing interaction is timed to ensure probe separability at crossing, while retaining probe--field correlations (Appendix~B).

The interaction generates a parity phase. Defining $|\pm\rangle=(|L\rangle\pm|R\rangle)/\sqrt2$, the ideal phase component is
\begin{equation}
 |\psi_{\Phi_{\rm em}}\rangle_{AB}
 =\cos\frac{\Phi_{\rm em}}2|++\rangle_{AB}
 -i\sin\frac{\Phi_{\rm em}}2|--\rangle_{AB},
 \label{eq:intro_em_pair}
\end{equation}
where $\Phi_{\rm em}$ is the additional post-crossing phase. Alice uses a weaker dipole and closes it quickly; Bob uses a stronger dipole and closes it slowly inside. This asymmetry bounds the emission record while allowing $\Phi_{\rm em}=\pi/2$ and positive one-way entanglement distillation.

Before her crossing, Alice transfers her decoupled internal qubit to an exterior register using an auxiliary EPR pair and a local Bell measurement \cite{Bennett1993}. This early transfer commutes with Bob's subsequent field interaction: usable entanglement is established after his slower closure. The Bell record travels outward and distillation records travel inward, requiring no reply from Bob. Gravity defines the horizon, while the quantum electromagnetic field mediates the interaction; Appendix~B gives proof of separability upon crossing, radiation bound, and causal timing.

\textit{Gravitationally Induced Entanglement.}---We now consider gravity itself as the mediator between invariant macroscopic masses \cite{Bose2017, Marletto2017}. Using a similar local freely falling description, we propose a gedankenexperiment predicting that bipartite entanglement can be dynamically generated between spatial superpositions separated by an event horizon. 

By exploiting the finite propagation speed of the mediating field, the exterior probe continuously interacts with the retarded Liénard-Wiechert potentials of the interior probe's pre-crossing history. While a full demonstration requires explicit linearized-quantum-gravity evolution of the gravitationally dressed states \cite{Danielson2023, Danielson2022}, evaluating this classical retarded potential across the probes' quantum spatial superpositions gives a leading-order phase model. It predicts post-crossing probe entanglement, conditional on a separable reduced crossing state compatible with the pre-existing probe--field correlations.

Consider radially infalling probes A (exterior) and B (interior) near a Schwarzschild black hole ($R_s = 2GM/c^2$), modeled as localized macroscopic masses with branch masses $m_{\text{br}} \approx 10.9\,\mu$g.

If the probes' spatial separation is microscopic compared to the curvature radius ($d \ll R_s$), we can adopt a Local Freely Falling (LFF) frame. Here, their relative radial acceleration $\ddot{d}$ balances background tidal stretching against mutual Newtonian attraction:
\begin{equation} 
\ddot{d} \approx \frac{c^2}{R_s^2}d - \frac{2Gm_{\text{br}}}{d^2}. 
\end{equation} 
For a supermassive black hole of mass $M = 10^8 M_\odot$ (the mass of Gargantua), setting $\ddot{d} = 0$ yields a stationary tidal equilibrium at $d_{\text{eq}} \approx 114\,\mu$m, establishing a causal window $\tau_{\max} \approx d_{\text{eq}}/c$ that scales as $M^{2/3}$. This validates the LFF assumption, as $114\,\mu\text{m} \ll R_s$. Over this bounded causal window ($\tau_{\max} \approx 0.38$~ps), radial and tangential tidal drifts remain on the order of the Planck length, $\mathcal{O}(10^{-35})$~m.

\begin{figure}[t]
\centering
\begin{tikzpicture}[scale=1.15, >=Stealth]
    \fill[gray!8] (-1.2, -1.2) -- (3.2, 3.2) -- (-1.2, 3.2) -- cycle; 
    \draw[->, black!60, thick] (-1.2, -2) -- (-1.2, 3.2) node[above, black] { $ct$ };
    \draw[->, black!60, thick] (-1.2, -2) -- (4.2, -2) node[right, black] { $z$ }; 
    \node[anchor=center, black!40, font=\scriptsize\sffamily\bfseries, align=center] at (1.2, 2.5) {BLACK HOLE\\INTERIOR};
    \node[anchor=center, black!40, font=\scriptsize\sffamily\bfseries, align=center] at (3.8, 2.5) {EXTERIOR\\SPACETIME}; 
    \draw[dashed, red!85!black, ultra thick] (-1.2, -1.2) -- (3.2, 3.2) node[pos=0.60, sloped, above=3pt, fill=gray!8, inner sep=1pt, font=\tiny\bfseries, text=red!85!black] {Horizon ($ct=z$)}; 
    \draw[dotted, green!60!black, semithick] (-1.1, -0.6) -- (4.0, -0.4) node[right, font=\tiny, black] {$\Sigma$};
    \draw[dotted, green!60!black, semithick] (-1.1, 0.7) -- (4.0, 0.9) node[right, font=\tiny, black] {$\Sigma$};
    \draw[dotted, green!60!black, semithick] (-1.1, 1.8) -- (4.0, 2.0) node[right, font=\tiny, black] {$\Sigma$}; 
    \draw[orange!85!black, ultra thick, decoration={markings, mark=at position 0.35 with {\arrow{>}}}, postaction={decorate}] (0, -2) -- (0, 0); 
    \draw[orange!85!black, thick, dashed] (0, 0) -- (0, 3.0) node[above, orange!85!black, font=\tiny\bfseries] {Probe B}; 
    \draw[blue!75!black, ultra thick] (1.5, -2) -- (1.5, 0.7);
    \draw[blue!75!black, ultra thick, decoration={markings, mark=at position 0.80 with {\arrow{>}}}, postaction={decorate}] (1.5, 0.7) .. controls (1.500, 1.055) and (2.207, 1.759) .. (2.719, 2.400);
    \draw[blue!75!black, thick, dashed] (2.719, 2.400) .. controls (2.929, 2.663) and (3.107, 2.881) .. (3.190, 3.000) node[right, blue!75!black, font=\tiny\bfseries, xshift=2pt] {Probe A}; 
    \draw[dashed, purple!85!black, ultra thick, ->] (0, -1.0) -- (1.5, 0.7) node[pos=0.55, above=3pt, sloped, black!85, font=\tiny] {Retarded Signal}; 
    \filldraw[orange!85!black] (0, -1.0) circle (2.2pt);
    \node[black, font=\tiny\bfseries, anchor=north west, align=left] at (0.15, -1.0) {Emission\\($t_{\mathrm{ret}} \le 0$)};
    \filldraw[black] (0, 0) circle (2.2pt);
    \node[anchor=south east, black, font=\tiny\bfseries, align=right, xshift=-2pt, yshift=2pt] at (0, 0) {Crossing\\($t=0$)};
    \filldraw[blue!75!black] (1.5, 0.7) circle (2.2pt);
    \node[anchor=north west, black, font=\tiny\bfseries, align=left, xshift=3pt, yshift=-1pt] at (1.5, 0.65) {Braking Begins\\($t = \tau_1$)}; 
    \draw[decorate, decoration={snake, amplitude=1.2pt, segment length=3.5pt}, orange!90!black, semithick] (1.86, 1.10) -- (2.26, 1.30);
    \draw[decorate, decoration={snake, amplitude=1.2pt, segment length=3.5pt}, orange!90!black, semithick] (2.20, 1.60) -- (2.60, 1.80);
    \draw[decorate, decoration={snake, amplitude=1.2pt, segment length=3.5pt}, orange!90!black, semithick] (2.58, 2.10) -- (2.98, 2.30); 
    \node[orange!95!black, font=\tiny\bfseries, align=left, anchor=west] at (2.70, 1.65) {Soft-Graviton\\Bremsstrahlung\\$\Gamma > 0$}; 
    \node[draw=orange!80!black, fill=white, rounded corners, semithick, align=left, font=\scriptsize] at (1.5, -2.9) {
        \textbf{Causal Decoherence Bound:} \\
        $\Gamma \ge \frac{729}{160\pi} \Phi_{\rm hold}$ (sharp) \\
        $\Gamma > \frac{6}{11\pi} \Phi$ (rigorous)
    }; 
\end{tikzpicture}
\caption{Spacetime diagram. Retarded signals (purple) propagate from Probe B's pre-crossing history. To escape engulfment, Probe A undergoes rapid, non-inertial radial deceleration, triggering soft-graviton emission (orange) that exponentially degrades entanglement.}
\label{fig:lif_spacetime}
\end{figure}
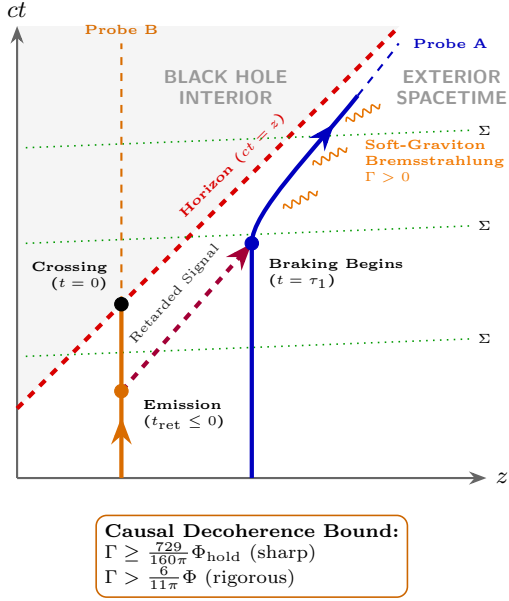

With $t = 0$ defined as the moment Probe B crosses the event horizon, we prepare the localized masses as tangential spatial superpositions (with branch separation $\Delta x$) prior to crossing. Within the pure, equal-amplitude phase-only model, initial separability is equivalent to the condition $\Theta(0) = \theta_{LL} + \theta_{RR} - \theta_{LR} - \theta_{RL} = 0 \pmod{2\pi}$ (equivalently $\Phi(0) = 0 \pmod{\pi}$ for the symmetric convention), which we impose by a pre-crossing phase echo. Reduced-state separability additionally requires compatible field overlaps: the correlated crossing state cannot be the pure product state of this phase-only approximation. The spatial branches sit at identically paired radial depths, rendering background classical potentials as trivial global phases that factor out. Entanglement is generated solely by the mutual retarded Liénard-Wiechert potentials originating from both probes' causal histories prior to crossing ($t_{\text{ret}} \le 0$), thereby respecting local causality (Fig.~\ref{fig:lif_spacetime}).

\begin{figure}[t]
\centering
\begin{tikzpicture}[x=1.0cm, y=1.0cm, >={Stealth[length=2mm, width=1.5mm]}]
\shade[top color=white, bottom color=red!5] (-3.2, 0.8) rectangle (3.2, -0.5); 
\draw[dashed, red!80!black, ultra thick] (-3.2, -0.5) -- (3.2, -0.5) node[midway, below=4pt, font=\small\bfseries] {Black Hole Event Horizon ($r \to R_s$)};
\draw[->, thick, purple!80!black] (-2.7, 3.5) -- (-2.7, 0.5) node[midway, left, align=center, font=\scriptsize\bfseries] {Gravity \\ Gradient};
\draw[dashed, purple!50, semithick] (-3.0, 3.5) -- (1.8, 3.5) node[right, font=\scriptsize] {Surface $r_A$}; 
\draw[dashed, purple!50, semithick] (-3.0, 0.5) -- (1.8, 0.5) node[right, font=\scriptsize] {Surface $r_B$};
\def\dx{2.2}   \def\deq{3.0}  
\coordinate (AL) at (-\dx/2, \deq+0.5); 
\coordinate (AR) at (\dx/2, \deq+0.5); 
\coordinate (BL) at (-\dx/2, 0.5); 
\coordinate (BR) at (\dx/2, 0.5);
\draw[ultra thick, blue!60, <->, shorten <=6pt, shorten >=6pt] (AL) -- (BL) node[midway, left, font=\small, black] { $d_{\text{eq}}$ }; 
\draw[ultra thick, blue!60, <->, shorten <=6pt, shorten >=6pt] (AR) -- (BR) node[midway, right, font=\small, black] { $d_{\text{eq}}$ };
\draw[thick, orange!90!black, densely dashed, <->, shorten <=6pt, shorten >=6pt] (AL) -- (BR) node[near start, sloped, above, font=\footnotesize, black] { $r_{\times}$ }; 
\draw[thick, orange!90!black, densely dashed, <->, shorten <=6pt, shorten >=6pt] (AR) -- (BL) node[near start, sloped, above, font=\footnotesize, black] { $r_{\times}$ };
\draw[<->, black!60, semithick] (-\dx/2, \deq+1.0) -- (\dx/2, \deq+1.0) node[midway, above, font=\footnotesize, text=black] {Tangential Width $\Delta x$};
\filldraw[ball color=gray!20, draw=black!80, thick] (AL) circle (4.5pt) node[above left, font=\small] { $|L\rangle_A$ }; 
\filldraw[ball color=gray!20, draw=black!80, thick] (AR) circle (4.5pt) node[above right, font=\small] { $|R\rangle_A$ }; 
\filldraw[ball color=gray!20, draw=black!80, thick] (BL) circle (4.5pt) node[below left, font=\small] { $|L\rangle_B$ }; 
\filldraw[ball color=gray!20, draw=black!80, thick] (BR) circle (4.5pt) node[below right, font=\small] { $|R\rangle_B$ };
\node[draw, fill=white, rounded corners, semithick, align=left, font=\scriptsize] at (0, -2.1) { \textbf{Phase Accumulation Mechanics:} \\ Parallel states ($|LL\rangle, |RR\rangle$) acquire $\phi_{||} \propto 1/d_{\text{eq}}$ \\ Crossed states ($|LR\rangle, |RL\rangle$) acquire $\phi_{\times} \propto 1/r_{\times}$ \\ $r_{\times} = \sqrt{d_{\text{eq}}^2 + \Delta x^2} \implies$ Parity Phase $\Phi = \phi_{||} - \phi_{\times}$ }; 
\end{tikzpicture}
\caption{Geometry of the Gravitational Parity-Phase gate. Tangential spatial superpositions implement the parity-phase geometry while avoiding unequal radial depths and asynchronous branch crossings.}
\label{fig:tangential_geometry}
\end{figure}
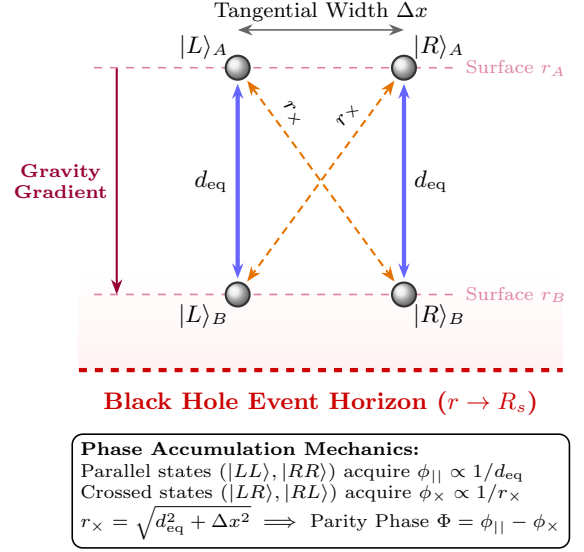

This symmetric tangential geometry enacts a gravitational parity-phase gate (Fig.~\ref{fig:tangential_geometry}). The distance between cross-aligned branches forms the hypotenuse $r_{\times} = \sqrt{d_{\text{eq}}^2 + \Delta x^2}$. Expanding the relative Newtonian potential yields the dynamical entangling phase: 
\begin{equation} 
\Phi \approx \frac{1}{2} \frac{G m_{\text{br}}^2}{\hbar d_{\text{eq}}} \left(\frac{\Delta x}{d_{\text{eq}}}\right)^2 \tau_1. 
\end{equation} 
Collinear setups accumulate phase faster but suffer asynchronous horizon-crossings and branch-dependent redshift phases. Our symmetric tangential geometry is sufficient for the present construction, bounding the phase within the LFF causal domain to $\Phi < \frac{1}{2} \Phi_{\text{max}} \left(\frac{\Delta x}{d_{\text{eq}}}\right)^2$. For $m_{\text{br}} \approx 10.9\,\mu$g, $\Phi_{\text{max}} = G m_{\text{br}}^2 / \hbar c \approx 0.25$~rad, accommodating weak entanglement regimes prior to spatial recombination. The tangential interaction generates the parity-phase state: 
\begin{equation} 
|\Psi(\Phi)\rangle = \frac{1}{2}\left( |LL\rangle + e^{-i\Phi}|LR\rangle + e^{-i\Phi}|RL\rangle + |RR\rangle \right). 
\end{equation} 
Applying local optical delays to the $|R\rangle$ branches identically maps the unitary to a Controlled-Phase gate $C\text{Phase}(2\Phi)$. This symmetrical gravitational parity-phase gate conditionally enables cross-horizon entanglement swapping via teleportation of Alice's probe \cite{Bennett1993}, leaving a localized mass inside the black hole entangled with a system in the exterior universe.

\textit{Causally Bounded Entanglement Extraction.}---Avoiding horizon engulfment introduces a strict sequential constraint: radial braking follows the entangling interval $\tau_1$ and must keep the probe outside the advancing horizon. Without braking, the causal window closes at $\tau_{\max} \approx d_{\text{eq}}/c$. Accelerating macroscopic stress-energy generates non-inertial vacuum bremsstrahlung (the Danielson-Satishchandran-Wald effect \cite{Danielson2023, Danielson2022}). 

To preserve the spatial superposition, decelerating momentum must be absorbed into a symmetrically centered apparatus ($x_{\text{app}} = 0$) to avoid which-path leakage. We restrict to a leading-order extraction model with monotonic braking and centralized controllers that do not destructively cancel the probe's cross-quadrupole radiation ($\Delta Q_{xz}(t) = 3m_{\mathrm{rad}}\Delta x\,z_p(t)$, $m_{\mathrm{rad}}\ge m_{\mathrm{br}}$). 

Within linearized quantum gravity in the leading quadrupole approximation \cite{Danielson2023, Danielson2022}, vacuum dephasing is governed by: 
\begin{equation} 
\Gamma = \frac{G}{90 \pi \hbar c^5} \int_0^\infty \omega \sum_{i,j} \left| \Delta \tilde{\ddot{Q}}_{ij}(\omega) \right|^2 \, d\omega. 
\end{equation} 
Defining the trace-free tensor $Q_{ij} = \frac{1}{c^2} \int T^{00} (3x_i x_j - r^2 \delta_{ij}) \, d^3x$ establishes the cross-term sum $\sum |\Delta \tilde{\ddot{Q}}_{ij}|^2 = 18 m_{\text{rad}}^2 \Delta x^2 |\tilde{\ddot{z}}_p(\omega)|^2$. Parameterizing the deceleration trajectory over $s = t/\tau_2$ such that $\ddot z_p(t)=+\frac{c}{\tau_2}a(s)$ with $\int_0^1 a(s) ds = 1$ yields:
\begin{equation}
\Gamma = \frac{Gm_{\mathrm{rad}}^2\Delta x^2}{5\pi\hbar c^3\tau_2^2}\,I[a], \quad I[a] = \int_0^\infty k |\tilde{a}(k)|^2 \, dk.
\end{equation}
To avoid horizon engulfment, stopping displacement $\Delta z_{\mathrm{stop}} = c\tau_2 \mu$, where $\mu=\int_0^1s\,a(s)\,ds$, dictates $\tau_2\le (\tau_{\max}-\tau_1)/\mu$. 

Defining the variational constant $C_* = \inf (\mu^2 I[a])$ over admissible profiles subject to physical monotonic-braking constraints (see Appendix A), we establish a sharp infimum over this non-negative profile class: $C_* = 27/16$.

The exact Newtonian parity phase accumulated before braking is
\begin{equation}
\Phi_{\rm hold}=
\frac{Gm_{\mathrm{br}}^2}{\hbar c}\,
x\left(1-\frac1{\sqrt{1+r^2}}\right).
\end{equation}
Using $\tau_{\max}=d_{\mathrm{eq}}/c$ and defining
\begin{equation}
x=\frac{\tau_1}{\tau_{\max}},
\qquad
r=\frac{\Delta x}{d_{\mathrm{eq}}},
\end{equation}
the causal stopping constraint and $C_*=27/16$ yield
\begin{equation}
\Gamma \ge
\frac{C_*}{5\pi}
\left(\frac{m_{\mathrm{rad}}}{m_{\mathrm{br}}}\right)^2
\frac{r^2}{1-1/\sqrt{1+r^2}}\,
\frac{\Phi_{\rm hold}}{x(1-x)^2}.
\end{equation}
The temporal partition maximizes at $x=1/3$ (yielding $4/27$), and the geometric factor bounds to $2$ as $r \to 0$. Taking the least-radiating member of the given no-cancellation model, $m_{\mathrm{rad}}=m_{\mathrm{br}}$, and applying the sharp temporal and geometric infima yields
\begin{equation} 
\Gamma \ge \frac{729}{160\pi} \Phi_{\rm hold}. 
\end{equation} 
The coefficient is approached for $x=1/3$ and $r\to0$; no finite nonzero $r$ saturates every strict inequality simultaneously. Because the optimized trajectory natively approaches relativistic regimes ($v \to c$) where higher-order corrections to the Newtonian quadrupole formula become structurally important \cite{Blanchet2014}, this represents a sharp infimum specifically governed by the sequential, non-negative braking, no-cancellation, leading-order quadrupole model rather than an absolute QFT bound. The formal endpoint $v=c$ is not a finite-energy massive trajectory.

Braking also accumulates phase. Minimizing the coupled radiation-to-phase penalty gives the global sharp bound proved in Appendix~A (where $\Phi = \Phi_{\rm hold} + \Phi_{\rm brake}$ and $\Gamma\ge C_{\rm sharp}\Phi$ for $C_{\rm sharp}\simeq0.17363$):
\begin{equation}
\Gamma > \frac{6}{11\pi}\Phi
\label{eq:total_phase_bound}
\end{equation}
within the same leading-order model.

Radial extraction triggers DSW bremsstrahlung, imposing an anisotropic $Z$-basis phase-damping channel where the spatial off-diagonal coherence decays as $D = \exp(-\Gamma)$. Including the braking phase, the Horodecki criterion \cite{Horodecki1995} bounds the maximum CHSH expectation value by:
\begin{equation} 
S_{\text{true}}(\Phi) \le 2\sqrt{\sin^2\Phi + \exp(-2\Gamma)}.
\end{equation} 
Equation~\eqref{eq:total_phase_bound} gives $D < \exp[-6\Phi/(11\pi)]$. Under standard phase-damping, the error probability is $p=(1-D)/2$.

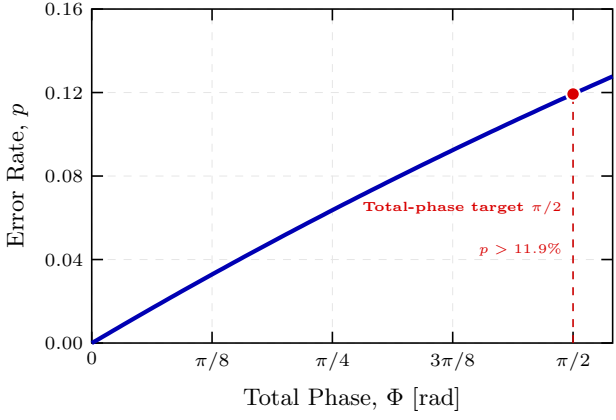
\begin{figure}[t]
\centering
\begin{tikzpicture}
\begin{axis}[
    width=0.98\columnwidth,
    height=6.0cm,
    xlabel={Total Phase, $\Phi$ [rad]},
    ylabel={Error Rate, $p$},
    xmin=0, xmax=1.7,
    ymin=0.0, ymax=0.16,
    ytick={0.0, 0.04, 0.08, 0.12, 0.16},
    yticklabels={$0.00$, $0.04$, $0.08$, $0.12$, $0.16$},
    xtick={0, 0.3927, 0.7854, 1.178, 1.5708},
    xticklabels={$0$, $\pi/8$, $\pi/4$, $3\pi/8$, $\pi/2$},
    grid=both,
    grid style={line width=.1pt, draw=gray!15},
    major grid style={line width=.2pt, draw=gray!20, dashed},
    axis lines=box,
    tick align=inside,
    tick style={black, semithick},
    tick label style={font=\scriptsize},
    label style={font=\small},
    axis line style={black, semithick},
    clip=false
]
    \addplot[domain=0:1.7, samples=100, blue!70!black, ultra thick] {0.5 * (1 - exp(-0.1736235 * x))};
    \addplot[dashed, red!80!black, semithick] coordinates {(1.5708, 0) (1.5708, 0.1193)};
    \node[circle, fill=red!85!black, draw=white, inner sep=1.8pt, thick] at (axis cs:1.5708, 0.1193) {};
    \node[anchor=east, red!85!black, font=\tiny\bfseries, xshift=-1pt] at (axis cs:1.5708, 0.065) {Total-phase target $\pi/2$};
    \node[anchor=east, red!85!black, font=\tiny\bfseries, xshift=-1pt] at (axis cs:1.5708, 0.045) {$p > 11.9\%$};
\end{axis}
\end{tikzpicture}
\caption{Rigorous lower bound on radial phase-damping error versus total Newtonian phase through braking. For $\Phi=\pi/2$, $p > 11.9\%$.}
\label{fig:extraction_error}
\end{figure}

\begin{figure}[t]
\centering
\begin{tikzpicture}
\begin{axis}[
    width=0.94\columnwidth,
    height=5.0cm,
    scale only axis=false,
    xlabel={Total Phase, $\Phi$ [rad]},
    ylabel={$S_{\mathrm{true}}(\Phi)$},
    xmin=0, xmax=1.7,
    ymin=0.0, ymax=3.0,
    ytick={0.0, 1.0, 2.0, 2.8284},
    yticklabels={$0$, $1.0$, $2.0$ (LHV), $2\sqrt{2}$},
    xtick={0, 0.3927, 0.7854, 1.178, 1.5708},
    xticklabels={$0$, $\pi/8$, $\pi/4$, $3\pi/8$, $\pi/2$},
    grid=both,
    grid style={line width=.1pt, draw=gray!15},
    major grid style={line width=.2pt, draw=gray!20, dashed},
    axis lines=box,
    tick align=inside,
    tick style={black, semithick},
    tick label style={font=\scriptsize},
    label style={font=\small},
    legend style={at={(0.03,0.97)}, anchor=north west, font=\fontsize{6}{7}\selectfont, draw=black!15},
    axis line style={black, semithick}
]
    \fill[gray!6, label=none] (axis cs:0, 0.0) rectangle (axis cs:1.7, 2.0);
    \addplot[domain=0:1.7, samples=100, blue!70!black, thick, dashed] {2 * sqrt(sin(deg(x))^2 + 1)};
    \addlegendentry{Ideal}
    \addplot[domain=0:1.7, samples=100, red!80!black, ultra thick] {2 * sqrt(sin(deg(x))^2 + exp(-0.347247 * x))};
    \addlegendentry{Bound}
    \addplot[dashed, red!80!black, semithick, forget plot] coordinates {(1.5708, 0) (1.5708, 2.514)};
    \node[circle, fill=red!85!black, draw=white, inner sep=1.8pt, thick] at (axis cs:1.5708, 2.514) {};
\end{axis}
\end{tikzpicture}
\caption{CHSH envelope using the rigorous total-phase dephasing bound. The model does not exclude Bell violation after radial braking.}
\label{fig:chsh_suppression}
\end{figure}
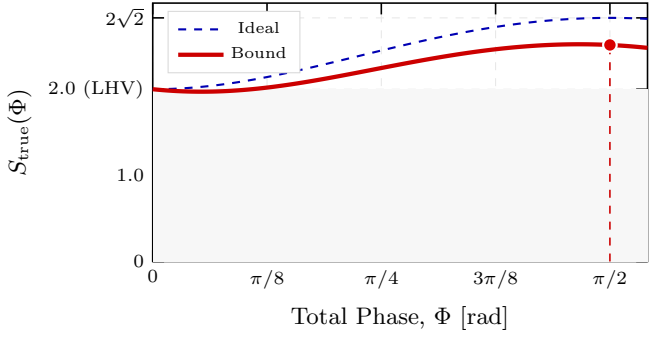

Crucially, because the event horizon prohibits two-way Local Operations and Classical Communication (LOCC) from the interior, purification must rely on one-way communication. For $\Phi=\pi/2$, the bound gives $\Gamma > 0.272$, $D < 0.762$, and $p > 11.9\%$ (Figs.~\ref{fig:extraction_error} and \ref{fig:chsh_suppression}). This rigorous bound excludes neither Bell violation nor one-way distillation \cite{Devetak2005}. Furthermore, evaluating the two-qubit symmetric-extension condition ($\operatorname{Tr}\rho_{AB}^2 > \operatorname{Tr}\rho_B^2$) dictates that distillation requires a residual coherence $D > |\cos\Phi|$. Combined with the radial dephasing bound $D \le \exp(-C_{\rm sharp}\Phi)$, this strictly precludes distillation for any total phase $\Phi \lesssim 0.34$~rad.

At higher phases, extraction faces severe mechanical constraints from rapid acceleration and acoustic shattering of solid probes. Optical probes bypass shattering but exchange macroscopic radial momentum $p_z \sim 2E/c$ with a steering apparatus upon reversal ($E$ being the field energy). Although a centered apparatus absorbing identical linear momentum prevents translational distinguishability, the tangential separation $\Delta x$ imparts opposite angular impulses, $J_\pm = \mp p_z\Delta x/2$. While such accessible torsional records could theoretically be erased, coupling to unmonitored thermal modes---or an infinitesimal fractional loss ($\ell \sim 10^{-28}$) from the $N \sim 10^{27}$ optical photons into the environment---exponentially suppresses branch overlap ($D_{\rm loss} = e^{-\ell N}$). These inaccessible records irreversibly drive the spatial coherence to zero.

\textit{Geometric Duality: Harvesting vs. Escape.}---While radial escape enforces a nonzero lower bound on dephasing within the given no-cancellation model, a distinct tangential harvesting protocol can bypass this dephasing limit as well as acoustic shattering. Under the assumed coherent control, Alice and Bob use macroscopic optical fields ($E_{\text{br}} \approx 1$~GJ $\implies m_{\text{br}} \approx 0.5 m_p$, where $m_p=\sqrt{\hbar c/G}$ denotes the Planck mass) confined strictly by electromagnetic potentials. The total energy within a $200\,\mu\text{m}^3$ trap has a nominal field scale below, but not parametrically below, the critical Schwinger scale ($1.3 \times 10^{18}$ V/m), meaning strong-field QED effects must be included in any quantitative controller analysis. 

At $t = \tau_1$, Alice begins tangential harvesting, culminating in a joint Bell-state measurement (BSM) between Probe A's decoded qubit and an auxiliary photon $C$. She switches off the optical traps, transferring macroscopic recoil momentum $\mp E_{\text{br}}/c$ strictly to the trapping apparatus. Because the left and right optical pulses and their fields trace exact mirror images ($x_R(t) = -x_L(t)$), the total stress-energy tensor ($T^{\mu\nu}_{\text{total}}$) maintains strict geometric symmetry. Under the trace-free diagonal definition $Q_{xx} = \frac{1}{c^2} \int T^{00}_{\text{total}} (2x^2 - y^2 - z^2) \, d^3x$, the spatial coordinates are squared, enforcing $Q_{xx}^{(R)}(t) = Q_{xx}^{(L)}(t)$ globally. Given transverse axisymmetry, off-diagonal components identically vanish. While operating on a highly relativistic light-crossing timescale ($\omega \Delta x / c \sim 1$) prevents strict multipole truncation, the displayed leading diagonal quadrupole distinguishability cancels under the assumed full-apparatus symmetry. Distinguishability from higher multipoles or departures from the assumed off-diagonal cancellation acts as a controller condition captured by the coherence multiplier $\eta$ (introduced below); we do not infer equality of the full radiation states from mirror symmetry alone when $\omega \Delta x / c$ is of order one.

Routing recombining macroscopic fluids via solid-state barriers transfers a transverse impulse $p_x$, requiring a zero-point spatial uncertainty $\sigma_X \le \hbar / (2p_x) \approx l_p$ for passive overlap. We instead model the central mixing apparatus as an effective operation using intense optical standing waves via Euler-Heisenberg vacuum nonlinearity \cite{Euler1936, Schwinger1951}. Letting the transferred recoil couple Alice's spatial branches to the coherent apparatus states $|G_L\rangle$ and $|G_R\rangle$, the bipartite state evolves to
\begin{equation}
\begin{split}
|\Psi\rangle \to \frac{1}{2} \Big[ &|L_B\rangle(|L_A G_L\rangle + e^{-i\Phi}|R_A G_R\rangle) \\
&+ |R_B\rangle(e^{-i\Phi}|L_A G_L\rangle + |R_A G_R\rangle) \Big].
\end{split}
\end{equation}
We treat near-critical confinement and effective operations as theoretical assumptions within the allowable bounds of unitary quantum mechanics and classical causal geometry, rather than taking the experiments of Refs.~\cite{Marquet2024, Sun2014} as demonstrations at the 1~GJ or sub-picosecond scale.

Under the phenomenological assumptions
\begin{equation}
\sigma_X\ge l_p,
\qquad
\sigma_X\sigma_P=\frac{\hbar}{2},
\end{equation}
a minimum-uncertainty Gaussian controller satisfies
\begin{equation}
\sigma_P\le\frac{\hbar}{2l_p}
=\frac12m_pc.
\end{equation}

We adopt $m_{\mathrm{br}}\simeq m_p/2$ as a reference design point. It yields a recoil $p_x\simeq m_pc/2$, equal to the largest controller momentum uncertainty permitted by the phenomenological minimum-length, minimum-uncertainty Gaussian model. Consequently, when $\sigma_P=m_pc/2$, the passive Gaussian overlap is
\begin{equation}
\mathcal V_G
=\exp\!\left[-\frac{p_x^2}{2\sigma_P^2}\right]
\simeq e^{-1/2},
\end{equation}
where the complex inner product dictates $\langle G_L|G_R\rangle = \mathcal{V}_G e^{i\chi}$, defining the relative apparatus phase $\chi$. This equality defines a convenient passive-overlap benchmark; it is not a fundamental mass bound, because ideal active recoil erasure remains algebraically possible for smaller overlaps, including $\mathcal V_G\to0$.

The passive-overlap benchmark does not itself preserve or destroy the logical state. Purity is restored by measuring the recoil controller coherently in the erasure basis
\begin{equation}
|e_\pm\rangle= \frac{|G_L\rangle\pm e^{-i\chi}|G_R\rangle} {\sqrt{2(1\pm\mathcal V_G)}}, \qquad \langle G_L|G_R\rangle=\mathcal V_Ge^{i\chi},
\end{equation}
obtaining a binary recoil-erasure outcome. When the controller records only Alice's branch, the logical Kraus operators are
\begin{equation}
K_+=\sqrt{\frac{1+\mathcal V_G}{2}}P(\chi),\qquad K_-=\sqrt{\frac{1-\mathcal V_G}{2}}ZP(\chi),
\end{equation}
where $P(\chi)=\operatorname{diag}(1,e^{i\chi})$. Because $K_+ \propto P(\chi)$ and $K_- \propto ZP(\chi)$, Alice applies a conditional logical $Z$ feed-forward for the minus outcome. After this feed-forward, both effective outcomes are proportional to $P(\chi) = \operatorname{diag}(1,e^{i\chi})$. This measurement thus acts as a local phase rotation on Alice's qubit that allows the known apparatus phase $\chi$ to be removed by $P(-\chi)$, leaving the bipartite gravitational phase $\Phi$ structurally intact. Introducing $\eta$ as the coherence multiplier from unmeasured branch-dependent leakage, ideal purity recovery assumes $|\eta|=1$ or a fully monitored correctable environment. Since $K_\pm^\dagger K_\pm \propto I$, both outcomes theoretically preserve purity under these exact conditions.

The controller erasure is assumed to be implemented separately, within a local readout duration $\tau_{\mathrm{read}}$; the optical decoding below does not by itself measure the controller. Coherent $L/R$ mode mixing must precede any unmonitored spatially resolving photon loss. We explicitly encode the branches as coherent states, $|L\rangle_A = |\alpha\rangle_L|0\rangle_R$ and $|R\rangle_A = |0\rangle_L|\alpha\rangle_R$. We apply a coherent 50:50 mode transformation to define bright and dark modes:
\begin{equation}
c_b=\frac{c_L+c_R}{\sqrt2},
\qquad
c_d=\frac{c_L-c_R}{\sqrt2}.
\end{equation}
This maps the branches to $|\beta\rangle_b|\beta\rangle_d$ and $|\beta\rangle_b|-\beta\rangle_d$, respectively, where $\beta = \alpha/\sqrt{2}$. The bright factor is common. We subject the dark mode to an ideal driven-dissipative two-photon deflation channel \cite{Marquet2024} governed by
\begin{equation}
\frac{d\rho}{dt} = -\frac{i}{\hbar}[H_2(t), \rho] + \kappa_b \mathcal{D}[b]\rho,
\end{equation}
where $H_2 = \hbar g_2(c_d^2 b^\dagger + c_d^{\dagger 2} b) + i\hbar(\epsilon_d b^\dagger - \epsilon_d^* b)$ uses a lossy buffer $b$. In the buffer-elimination regime, the effective jump operator is $L_2(t) = \sqrt{\kappa_2}(c_d^2 - \beta(t)^2)$. Adiabatically reducing the effective drive $\beta(t)$ to zero coherently maps the even and odd macroscopic cat components to the microscopic single-rail qubit in the $\{|0\rangle, |1\rangle\}$ manifold of the dark mode.

Continuous quantum-jump tracking \cite{Sun2014} on the dark collective channel monitors single-photon leakage under an idealized gedanken unit-efficiency assumption. This explicitly distinguishes correctable collective $c_d$ loss from destructive spatially resolved $c_L$ or $c_R$ loss (assumed zero in the lossless ideal limit). A bright-channel jump is branch independent. While the dark cat amplitude $\beta$ remains large, a dark-channel jump approximately applies a known logical $Z$, which Alice tracks and corrects using the monitored jump count modulo two, within the large-cat approximation. However, as $\beta \to 0$ in the terminal deflation stage, the code approaches the single-rail basis where a jump is no longer a unitary Pauli operation. Therefore, any single-photon jump detected during terminal deflation and single-rail readout is heralded as a failed trial and discarded, retaining only the no-jump records. The known relative attenuation from the no-jump evolution is absorbed into the final Procrustean filter, reducing the overall yield.

Applying an explicit local isometry into a dual-rail encoding (e.g., $V|0\rangle_d = |1,0\rangle_{uv}$ and $V|1\rangle_d = |0,1\rangle_{uv}$, together with any known Hadamard arising from the cat-parity basis), and accounting for its duration within $\tau_{\mathrm{read}}$, Alice interferes this qubit with auxiliary photon $C$ on a 50:50 beamsplitter. Ordinary photodetection acts as a heralded partial BSM, yielding the target Bell outcome, supplemented by the separately obtained recoil-erasure outcome, before the causal window closes.

For the sequential entangle--recombine protocol, we reserve $\Delta x/(2c)$ for optical recombination and include the remaining local operations in $\tau_{\mathrm{read}}$. Define
\begin{equation}
r=\frac{\Delta x}{d_{\mathrm{eq}}},
\qquad
\delta=\frac{c\tau_{\mathrm{read}}}{d_{\mathrm{eq}}}.
\end{equation}
The causal condition yields the allowed domain $r < 2(1-\delta)$ and
\begin{equation}
\frac{c\tau_1}{d_{\mathrm{eq}}}
\le 1-\frac r2-\delta,
\end{equation}
giving the maximal hold phase
\begin{equation}
\begin{split}
\Phi_{\mathrm{op}}(m_{\mathrm{br}},\delta)
&=
\frac{Gm_{\mathrm{br}}^2}{\hbar c}
\max_{0<r<2(1-\delta)}\\
&\quad{}
\left(1-\frac1{\sqrt{1+r^2}}\right)
\left(1-\frac r2-\delta\right).
\end{split}
\end{equation}
For the $\delta \to 0$ supremum, define
\begin{equation}
f_\star=
\max_{0<r<2}
\left(1-\frac1{\sqrt{1+r^2}}\right)
\left(1-\frac r2\right),
\end{equation}
which evaluates to $0.147495884\ldots$. The maximum occurs at $r_\star=1.069515334\ldots$. Therefore, for $m_{\mathrm{br}}=m_p/2$,
\begin{equation}
\Phi_\star=\frac{f_\star}{4},
\end{equation}
yielding $0.036873971\ldots$. This is the $\delta\to0$ hold-phase supremum, not a bound on the output phase or a universal uncertainty bound. Recombination adds a calculable propagation phase, while the complete output also depends on the controller and readout map (Appendix~C). Because $\tau_{\max}$ scales as $M^{2/3}$, increasing the black-hole mass drives $\delta$ toward zero for fixed local readout duration, allowing the hold phase to approach this supremum.

\textit{Discussion and Conclusion.}---Our symmetrical gravitational parity-phase gate conditionally enables cross-horizon entanglement swapping via teleportation of Alice's probe \cite{Bennett1993}. Suppose a free-falling Alice holds Probe A, gravitationally entangled with interior Bob (Probe B). Alice also holds photon $C$ of an auxiliary EPR pair $(C,D)$; $D$ is held in a nearby exterior register, as in Appendix~B. Concentration and its inward success flag precede outward release of the output carrier, ensuring that Bob can receive the herald before encountering the singularity. 

To fit Alice's local harvesting operations within the causal window and multiplex the probabilistic concentration, Alice and Bob distribute a spatially multiplexed array of independent probe pairs laterally across the locally flat event horizon. To suppress N-body gravitational crosstalk, a $10$-meter spacing restricts unwanted four-branch phase invariants from neighboring pairs to a geometry-qualified $2 \times 10^{-16}$ to $4 \times 10^{-16}$~rad range, making pairwise crosstalk negligible at this benchmark.

Each Alice synchronously performs the heralded BSM and recoil-erasure protocols, broadcasting the classical measurement outcomes radially outward before crossing the null boundary. For equal-amplitude, phase-only readout, the nearby exterior register applies the corresponding Pauli and phase corrections to photon $D$, obtaining
\begin{equation}
|\psi\rangle_{DB}
=
\cos\frac{\Phi_{\rm out}}{2}|++\rangle_{DB}
+i\sin\frac{\Phi_{\rm out}}{2}|--\rangle_{DB}.
\end{equation}
Here $\Phi_{\rm out}$ includes all retained joint phase. For $0<\Phi_{\rm out}\le\pi/2$, the one-sided Procrustean filter \cite{Bennett1996}
\begin{equation}
F_D=
\tan\frac{\Phi_{\rm out}}{2}|+\rangle\langle+|
+|-\rangle\langle-|
\end{equation}
produces a maximally entangled pair with probability
\begin{equation}
P_{\mathrm{dist}}
=2\sin^2\frac{\Phi_{\rm out}}{2}
\simeq\frac{\Phi_{\rm out}^2}{2}.
\end{equation}
No communication from Bob is required. In this gravitational protocol, all corrections act outside; Bob receives the concentration-success flag to identify the heralded pairs. By properly scaling laboratory dimensions with the black hole mass, Bob retains sufficient proper time to act as a passive receiver before encountering the singularity. The benchmark $\Phi_{\rm out}=\Phi_\star$ and a heralded 50\% BSM yield give one successful pair per approximately 2942 trials, excluding additional losses and rejected readouts; further joint phase or record-dependent attenuation changes this yield. Notably, our semiclassical approach does not violate the prohibition on generating entanglement from separable states by LOCC; rather, the classical retarded potential is an effective low-energy description of an underlying quantum field, consistent with recent proposals for local entanglement mediated by a quantized gravitational field \cite{Nandi2024}. 

Recently, a higher-order entangling mechanism involving quantum matter propagation in a classical gravitational field has been proposed \cite{AzizHowl2025}. Its predicted contribution is much smaller than gravitational mediation in the low-mass, long-duration examples considered there, but the comparison is parameter-dependent. Its derivation and interpretation remain debated \cite{Marletto2025,Gundhi2026}. Our gravitational scheme instead addresses the leading quantum-gravitational interaction and the associated soft-graviton decoherence.

Our electromagnetic construction realises continued post-crossing interaction without requiring quantised gravity. Alice couples only to Bob's pre-crossing field, while Bob subsequently receives Alice's final field change. After Alice's early transfer and Bob's slow closure, their retained registers share distillable entanglement. The mediating field thus preserves correlations across the crossing, without invoking Hawking evaporation.

Overturning the deeply ingrained assumption that once a particle separable from another particle in the exterior universe crosses an event horizon it cannot dynamically become entangled with that particle—our thought experiment predicts, within a constrained leading-quadrupole model, an elegant feature of semiclassical gravity: Tangential symmetry in principle allows the transverse teleportation of cross-horizon gravitationally induced entanglement, while radial escape is ultimately precluded by macroscopic which-path leakage.

\textit{Data Availability Statement.}---No new data were created or analysed in this study. 

\textit{Conflict of Interest.}---The author declares no conflicts of interest.

\textit{Ethics Statement.}---Not applicable.

\textit{Acknowledgments.}---I am grateful to Tim Spiller and Bernard Kay for helpful discussions. And for invaluable support, I am grateful to Tim Spiller, Georgia Mortzou, Marco Lucamarini, and my co-founder at Path Not Taken, Mahmoud Ashmawy. This work was supported by EPSRC (Grant number EP/T001011/1). Google Gemini Deep Think and OpenAI Codex assisted with manuscript preparation, derivations, and numerical checks. I reviewed their contributions and take responsibility for the manuscript.

\appendix
\section{Appendix A: Extraction-Trajectory Bounds} \label{app:A}
We prove $C_*=\inf_{a\in\mathcal A}\mu^2I[a]=27/16$, where $\mu=\int_0^1sa(s)ds$ and $\mathcal A$ contains non-negative, normalized, finite-energy profiles supported in $[0,1]$. Because the DSW bremsstrahlung is proportional to $\int_0^\infty \omega |\tilde{\ddot{Q}}(\omega)|^2 \, d\omega$, the vacuum decoherence penalty natively imposes an $H^{1/2}$ fractional Sobolev energy norm in the time domain. Extend profiles by zero and set $\widetilde a(k)=\int a(s)e^{-iks}ds$, $|D|=(-\partial_s^2)^{1/2}$, and $\mathcal E(f,g)=\pi\int f|D|g$, so $I[f]=\mathcal E(f,f)$. Non-negativity enforces monotonic braking; allowing signed profiles would give $C_*=0$.

Let
\begin{equation}
q(s)=\sqrt{s(1-s)}\,\mathbf 1_{(0,1)}(s).
\end{equation}
For the zero extension,
\begin{equation}
|D|q=1,\qquad |D|(sq)=2s-\frac12, \qquad 0<s<1.
\end{equation}
These follow from $|D|=\mathcal H\partial_s$, using $\mathcal H[\sqrt{1-x^2}]=x$ and $\mathcal H[x\sqrt{1-x^2}]=x^2-\tfrac12$ on $(-1,1)$. Thus
\begin{equation}
a_0(s)=\frac{16}{\pi}\sqrt{s}(1-s)^{3/2}
\end{equation}
is normalized, has mean $\mu_0=3/8$, and satisfies
\begin{equation}
|D|a_0=\frac8\pi(3-4s),\qquad I[a_0]=12.
\end{equation}
Hence $C_*\le27/16$, also approached by smooth non-negative approximations.

For $0<\mu\le3/8$, set $L=8\mu/3$ and $a_L(s)=L^{-1}a_0(s/L)\,\mathbf 1_{(0,L)}(s)$, with $\ell_L(s)=\frac8{\pi L^2}\left(3-\frac{4s}{L}\right)$. Then $|D|a_L=\ell_L$ on $(0,L)$. For $s>L$, writing $y=2s/L-1>1$, direct evaluation yields the obstacle inequality $|D|a_L(s)-\ell_L(s) = \frac8{\pi L^2}\frac{(2y+1)(y-1)}{\sqrt{y^2-1}} \ge0$. Hence, for any admissible $a$ with the same normalization and mean, $\mathcal E(a,a_L) \ge \pi\int_0^1a(s)\ell_L(s)\,ds = I[a_L]$. 

Since
\begin{equation}
0\le I[a-a_L] =I[a]+I[a_L]-2\mathcal E(a,a_L) \le I[a]-I[a_L],
\end{equation}
we obtain $I[a]\ge I[a_L]=\frac{12}{L^2}$ and $\mu^2I[a]\ge\frac{27}{16}$.

For $3/8\le\mu\le5/8$, the admissible profile $a_\mu(s)=\sqrt{s(1-s)}\left[\frac{8(5-8\mu)}{\pi}+\frac{64(2\mu-1)}{\pi}s\right]$ has mean $\mu$ and affine $|D|a_\mu$. Thus $h=a-a_\mu$, with zero mass and first moment, obeys $\mathcal E(h,a_\mu)=0$, giving $I[a]\ge I[a_\mu]=256\mu^2-256\mu+72$. Since
\begin{equation}
\frac{d}{d\mu}\!\left[\mu^2I[a_\mu]\right] =16\mu(8\mu-3)^2\ge0.
\end{equation}
the minimum is again $27/16$.

For $\mu\ge5/8$, reflection $a(1-s)$ preserves $I$ and has mean $\nu=1-\mu\le3/8$, so $\mu^2I[a]\ge\nu^2I[a]\ge27/16$. This completes the sharp bound.

\textit{Including the Braking Phase.}---
Including braking in the same non-negative, no-cancellation, Newtonian-phase and leading-quadrupole model, write $d=d_{\rm eq}$, $y=ct/d$, $r=\Delta x/d$, $\alpha=Gm_{\rm br}^2/(\hbar c)$, and
\begin{equation}
\begin{gathered}
f\ge0,\quad\int_0^\infty f=1,\quad\int_0^\infty yf(y)\,dy=\mu\le1,\\
q_f(y)=1+\int_0^y(y-u)f(u)\,du.
\end{gathered}
\end{equation}
With the same Fourier convention and zero extension, require finite energy:
\begin{equation}
\begin{gathered}
I[f]=\int_0^\infty k|\widehat f(k)|^2dk=\pi\langle f,|D|f\rangle,\\
\mathcal P[f]=\int_0^\infty q_f(y)^{-3}dy.
\end{gathered}
\end{equation}
For a finite phase endpoint $Y$, the exact Newtonian model gives
\begin{align}
 \Phi_Y&=\alpha\int_0^Y\left[q_f^{-1}-(q_f^2+r^2)^{-1/2}\right]dy,\\
 \Gamma&=\frac{\alpha r^2}{5\pi}
 \left(\frac{m_{\rm rad}}{m_{\rm br}}\right)^2 I[f].
\end{align}
Since $0<q^{-1}-(q^2+r^2)^{-1/2}\le r^2/(2q^3)$,
\begin{equation}
 \frac{\Gamma}{\Phi_Y}\ge\frac{2}{5\pi}
 \left(\frac{m_{\rm rad}}{m_{\rm br}}\right)^2\frac{I[f]}{\mathcal P[f]}.
 \label{eq:sharp_reduction}
\end{equation}
For $f_L(y)=L^{-1}f(y/L)$, differentiation at $L=1$ gives
\begin{equation}
\begin{gathered}
\frac{dI[f_L]}{dL}=-2I[f],\\
\frac{d\mathcal P[f_L]}{dL}=-2\mathcal P[f]+3E_4[f],\quad
E_4[f]=\int_0^\infty q_f^{-4}dy.
\end{gathered}
\end{equation}
Thus $d(I/\mathcal P)/dL=-3IE_4/\mathcal P^2<0$: the mean constraint saturates at an optimum. The exact coefficient is
\begin{equation}
 C_{\rm sharp}=\frac{2\lambda_*}{5\pi},\qquad
 \lambda_* =\min_{f\ge0,\ \int f=\int yf=1}\frac{I[f]}{\mathcal P[f]}.
 \label{eq:sharp_constant}
\end{equation}
\textit{Existence and Support.}---
For mass-one profiles with mean at most one,
\begin{equation}
 \max(1,y)\le q_f(y)\le1+y,\qquad \frac12\le\mathcal P[f]\le\frac32.
\end{equation}
A minimizing sequence has bounded $I$ and, by $|\widehat f|\le1$ and a frequency split, bounded $L^2$ norm. Local $H^{1/2}$ compactness and $\int_R^\infty f\le1/R$ preserve mass one, with limiting mean at most one. Lower semicontinuity of $I$, dominated convergence of $\mathcal P$, and dilation give a mean-one minimizer.

For a minimizer $g$, $\lambda=I[g]/\mathcal P[g]$, and mass and mean multipliers $a,b$, stationarity of $I-\lambda\mathcal P$ gives
\begin{equation}
\begin{gathered}
W_g(u)=2\pi|D|g(u)+3\lambda V_g(u)+a+bu,\\
W_g\ge0,\quad gW_g=0,\\
V_g(u)=\int_u^\infty(y-u)q_g(y)^{-4}dy.
\end{gathered}
\label{eq:sharp_obstacle}
\end{equation}
These weak half-line conditions admit independent moment variations where $g>0$. Amplitude scaling gives $2I+3\lambda(\mathcal P-E_4)+a+b=0$; time dilation gives $b=3\lambda E_4$. Hence
\begin{equation}
 a=-5I[g],\qquad b=3\lambda E_4[g]>0.
 \label{eq:sharp_multipliers}
\end{equation}
Using the affine obstacle for $a_0$, set
\begin{equation}
\begin{gathered}
L_B=\frac{20I[g]}{3b}=\frac{20\mathcal P[g]}{9E_4[g]},\\
K=\frac{5I[g]L_B^2}{48},\quad B(u)=\frac K{L_B}a_0(u/L_B).
\end{gathered}
\end{equation}
Then $2\pi|D|B\ge5I[g]-bu$ is a supersolution since $V_g\ge0$. Testing against $(g-B)_+$ in the positive fractional Dirichlet form gives $g\le B$, without assuming connected support.

Convexity of $q_g$ makes $dy/dq$ non-increasing. Opposite monotonicities under the measure $q^{-4}dq$ give
\begin{equation}
 \frac{\mathcal P[g]}{E_4[g]}\le
 \frac{\int_1^\infty q^{-3}dq}{\int_1^\infty q^{-4}dq}=\frac32.
\end{equation}
An initial hold adds equally to both integrals, preserving the bound. Thus every global minimizer has
\begin{equation}
 \operatorname{supp} g\subset[0,R],\qquad R=\frac{10}{3}.
 \label{eq:sharp_support}
\end{equation}

\textit{Uniqueness and Certification.}---
The Dirichlet half-line Green kernel for $|D|$ is
\begin{equation}
 G(s,t)=\frac1\pi\log\frac{\sqrt s+\sqrt t}{|\sqrt s-\sqrt t|}.
\end{equation}
by the increasing-interval limit of Ref.~\cite{Cauchy}, Sec.~11.1. Energy Cauchy--Schwarz for $H_h(y)=\int_0^y(y-u)h(u)du$ gives
\begin{equation}
 |H_h(y)|^2\le\frac{4y^4}{9\pi^2}I[h].
 \label{eq:sharp_greenbound}
\end{equation}
Here $\int_0^1\!\int_0^1(1-s)(1-t)\log[(\sqrt s+\sqrt t)/|\sqrt s-\sqrt t|]dsdt=4/9$, with the two factors $1/\pi$ supplied by $G$ and $I$.

For admissible mean-one $g,g+h$ supported in $[0,R]$, $H_h=0$ beyond $R$. Taylor's theorem gives
\begin{align}
 0\le\mathcal P[g+h]-\mathcal P[g]-\mathcal P'[g]h&\le6K_RI[h],\\
 K_R&=\frac4{9\pi^2}\left(\frac15+\log R\right).
 \label{eq:sharp_convexity}
\end{align}
Hence $F_\lambda=I-\lambda\mathcal P$ is strongly convex on this fixed-support, fixed-moment class if $6\lambda K_R<1$.

Scaling $a_0$ to support $[0,8/3]$ gives mean one and $I=27/16$. Its convex trajectory obeys $q(y)\le1+5y/8$ on the support and $q(y)=y$ thereafter, so
\begin{equation}
\begin{gathered}
\mathcal P\ge\int_0^{8/3}(1+5y/8)^{-3}dy+\int_{8/3}^\infty y^{-3}dy\\
=\frac{97}{128},\qquad\lambda_*\le\frac{216}{97}.
\end{gathered}
\end{equation}
Since $1-6(216/97)K_R>0.155$, strong convexity and Eq.~\eqref{eq:sharp_support} establish global uniqueness.

For an admissible mean-one trial $g$ supported in $[0,R]$, put $\lambda_g=I[g]/\mathcal P[g]$ and define $W_g$ by Eqs.~\eqref{eq:sharp_obstacle}--\eqref{eq:sharp_multipliers}. If $6\lambda_gK_R<1$, $|W_g|\le\epsilon$ on its support, and $W_g\ge-\epsilon$ elsewhere in $[0,R]$, then
\begin{equation}
 \lambda_g-\frac{\epsilon^2R^2}{4(1-6\lambda_gK_R)}
 \le\lambda_*\le\lambda_g.
 \label{eq:sharp_certificate}
\end{equation}
Indeed, the zero-extension Dirichlet form gives
$I[h]\ge\int_0^R h(y)^2/y\,dy$, hence
$\|h\|_1\le R\sqrt{I[h]/2}$. For $h=g_*-g$, Eq.~\eqref{eq:sharp_convexity} yields
\begin{equation}
\begin{gathered}
\begin{aligned}
F_{\lambda_g}[g_*]&\ge-\epsilon R\sqrt{I[h]/2}\\
&\quad +(1-6\lambda_gK_R)I[h]\\
&\ge-\frac{\epsilon^2R^2}{8(1-6\lambda_gK_R)}.
\end{aligned}
\end{gathered}
\end{equation}
Using $\mathcal P[g_*]\ge1/2$ proves the global certificate.

\textit{Numerical Enclosure.}---
Use $g(y)=L^{-1}A(y/L)$, where $U_n$ are Chebyshev polynomials of the second kind and
\begin{align}
 A(s)&=\frac8\pi\sqrt{s(1-s)}\sum_{n=0}^{12}a_nU_n(2s-1),\\
 a_0&=1,\quad a_1=-\frac12-\frac12\sum_{n=2}^{12}(n+1)a_n,\notag\\
 \mu&=\frac12+\frac{a_1}{4},\quad L=\mu^{-1},\\
 I[g]&=8\mu^2\sum_{n=0}^{12}(n+1)a_n^2.
\end{align}
Mass and mean are one, and the upper-end square-root term vanishes. Table~\ref{tab:sharp_coeff} defines exact rational inputs.

\begin{table}[t]
\centering\footnotesize
\begin{tabular}{rl}
\hline
$n$&$a_n$\\\hline
2&$-0.03830912227197032$\\
3&$0.009568329174274214$\\
4&$-0.0013873367197673966$\\
5&$-0.00008369257199057345$\\
6&$0.00010359104216184947$\\
7&$-0.000021079243227111604$\\
8&$-0.0000025591683205266023$\\
9&$0.0000025454392731921688$\\
10&$-0.000000516975830903794$\\
11&$-0.00000007505900449435073$\\
12&$0.00000006649372631557963$\\
\hline
\end{tabular}
\caption{Exact rational coefficients of the degree-12 trial; $a_0,a_1$ follow from normalization and endpoint cancellation.}\label{tab:sharp_coeff}
\end{table}

Writing $z=2s-1$, the polynomial is
$1-z+\sum_{n\ge2}a_n[U_n(z)-(n+1)z]$. Its positivity follows from
\begin{equation}
 \sum_{n=2}^{12}|a_n|(n+1)\left[\frac{n(n+2)}3+1\right]<0.738.
\end{equation}
Here $U_n(1)=n+1$, and the derivative bound used above is
\[
 \max_{|z|\le1}|U_n'(z)|=\frac{n(n+1)(n+2)}3.
\]
The trial support satisfies $L<2.595<R$.

For the numerical certificate, set
\begin{equation}
 M=\frac{4\mu}{\pi}\left(2+2\sum_{n=2}^{12}(n+1)|a_n|\right)\ge\|g\|_\infty.
\end{equation}
Since $0\le q'\le1$, $q''=g$ and $q\ge1$, the second derivatives of $q^{-3}$, $q^{-4}$ and $yq^{-4}$ on $[0,L]$ are bounded respectively by
\begin{equation}
\begin{gathered}
B_P=12+3M,\quad B_E=20+4M,\\
B_J=L(20+4M)+8.
\end{gathered}
\end{equation}
Trapezoidal integration with $N$ equal panels bounds full and partial integral errors by $e_j=L^3B_j/(12N^2)$, $j=P,E,J$. Since $q(y)=y$ beyond $L$, the tails are exact:
\begin{equation}
\begin{gathered}
\mathcal P_{\rm tail}=\frac1{2L^2},\quad(E_4)_{\rm tail}=\frac1{3L^3},\\
V_{\rm tail}(u)=\frac1{2L^2}-\frac u{3L^3}.
\end{gathered}
\end{equation}
If $P_N,E_N$ are the numerical integrals, enclose the exact trial ratio by
$\lambda_- = I/(P_N+e_P)$ and $\lambda_+=I/(P_N-e_P)$.
For $\widehat\lambda=I/P_N$, let $e_\lambda=\max(\widehat\lambda-\lambda_-,\lambda_+-\widehat\lambda)$.

Evaluate the residual at 4097 equally spaced points on $[0,L]$. Its curvature is bounded by
\begin{equation}
\begin{gathered}
M_W=\frac{64}{L^4}\sum_{n=2}^{12}(n+1)|a_n|\,d_n+3\lambda_+,\\
d_n=\frac{(n-1)n(n+1)(n+2)(n+3)}{15},
\end{gathered}
\end{equation}
using the endpoint bound on $|U_n''|$. With grid spacing $\Delta=L/4096$, a conservative uniform residual bound is
\begin{align}
 \epsilon={}&\max_j|W_N(j\Delta)|+\frac{M_W\Delta^2}{8}\notag\\
 &+3e_\lambda[1+L(E_N+e_E)]\notag\\
 &+3\lambda_+(e_J+2Le_E).
\end{align}
For $u>L$,
\begin{equation}
\begin{gathered}
\begin{aligned}
W_g'(u)&=4\int_0^L\frac{g(s)}{(u-s)^3}ds\\
&\quad-\frac{\lambda_g}{u^3}+3\lambda_gE_4[g]>0,
\end{aligned}
\end{gathered}
\end{equation}
since $E_4\ge1/3$ and $L>1$. The $3/2$ endpoint vanishing makes $W_g$ continuous at $L$, extending the bound to $[0,R]$.

The calculation uses 50-digit decimal arithmetic, with $10^{-25}$ rounding allowances in errors and endpoints. Two half-angle reductions give an inverse-tangent series argument below $0.2$; 45 terms leave a remainder below $10^{-63}$. Rounded square roots, logarithms, and bounded degree-12 recurrences contribute well below the allowance. This is an analytic-error certificate, not a proof-assistant formalization.

At $N=65536$, $e_P<5.225\times10^{-9}$, $\epsilon<6.881\times10^{-6}$, and $1-6\lambda_+K_R>0.48269$. Equation~\eqref{eq:sharp_certificate}, with outward ratio bounds, gives
\begin{equation}
 \boxed{0.1736326330<C_{\rm sharp}<0.1736326346.}
\end{equation}
Thus
\begin{equation}
\begin{gathered}
\Gamma\ge C_{\rm sharp}\Phi>\frac6{11\pi}\Phi,\\
\Phi>0,
\end{gathered}
\end{equation}
where the least-radiating case has $m_{\rm rad}=m_{\rm br}$. No simple closed form for $C_{\rm sharp}$ is claimed.

\textit{Sharpness and Scope.}---
Equation~\eqref{eq:sharp_reduction} covers every finite endpoint. To approach equality, smoothly approximate the minimizer with a small initial hold and compression, retaining positive causal clearance; then take $r\to0$ and increasing finite endpoints. Zero-acceleration padding, or a positive late tail of vanishing mass, first moment and energy, recovers the full phase integral. Hence no larger coefficient holds in this formal class.

The limit has vanishing branch width and clearance, unbounded duration, and net velocity change $c$; it is not a finite-energy massive trajectory. Relativistic dynamics, finite curvature-scale duration, or controller restrictions require a different optimization. The coefficient concerns total Newtonian phase; $729/(160\pi)$ remains sharp for the initial hold alone.

\section{Appendix B: Electromagnetically Induced Entanglement Across an Event Horizon}
\label{app:charged}

\textit{Confined Dipole Superpositions.}---We retain the local freely falling geometry, with radial separation $d$ and $\tau=d/c$. Each neutral probe has two coherently labelled charge configurations with opposite tangential dipoles, $\hat{\mathbf p}_j=p_jf_j(t)Z_j\mathbf e_x$, where $Z|L\rangle=|L\rangle$, $Z|R\rangle=-|R\rangle$ and $|\pm\rangle=(|L\rangle\pm|R\rangle)/\sqrt2$. For a bound charge displaced by $\pm\Delta_j/2$ relative to a compensating distribution, $p_j=q_j\Delta_j/2$. Both branches have the same net charge; closing the displacement retains the internal qubit. Dipole qubits and switchable interactions have established precedents \cite{ChargeDeMille2002,ChargeYelin2006}; here we use a controlled polarization model, not the bare Coulomb interaction of isolated charges.

To specify a finite, conserved source, take
\begin{align}
 \mathbf P_j(\mathbf x,t)&=p_j f_j(t)Z_j\mathbf e_x u_j(\mathbf x-\mathbf x_j),\notag\\
 \rho_j^{\rm ch}&=-\boldsymbol\nabla\cdot\mathbf P_j,
 &\mathbf j_j&=\partial_t\mathbf P_j,
 \label{eq:charge_polarization}
\end{align}
with smooth, nonnegative, normalized spherical profiles $u_j$ of radii $a_j\ll d$. Common binding charge belongs to the apparatus. For the free interaction-picture electric field $E_x$, the coupling is
\begin{align}
 H_I&=-\sum_{j=A,B}p_j f_j(t)Z_j E_{x,u_j}(t),\notag\\
 E_{x,u_j}&=\int u_j E_x\,d^3x.
 \label{eq:charge_dipole_H}
\end{align}
Local polarization self-energies are branch-independent, and contact cross terms vanish for disjoint supports. We assume coherent confinement, nonrelativistic charge motion, and negligible residual center motion and gravitational branch corrections. The total apparatus size and duration satisfy $\mathcal L,cT\ll R_s$, with negligible backreaction; switching times exceed $a_j/c$.

\textit{Retarded Phase and Crossing Separability.}---For transverse point dipoles, define
\begin{equation}
 \mathcal K_\tau f(t)=f(t-\tau)+\tau\dot f(t-\tau)
                         +\tau^2\ddot f(t-\tau).
\end{equation}
Bob's sourced field at Alice is $E_{B,x}=-k_ep_BZ_B\mathcal K_\tau f_B/d^3$, with $k_e=(4\pi\epsilon_0)^{-1}$. The two directed phases and the parity phase are
\begin{align}
 \phi_{AB}(t)&=\frac{4k_ep_Ap_B}{\hbar d^3}
       \int_{-\infty}^t f_A(s)\mathcal K_\tau f_B(s)\,ds,\notag\\
 \phi_{BA}&=\phi_{AB}\big|_{A\leftrightarrow B},\qquad
 \Phi=\frac{\phi_{AB}+\phi_{BA}}4.
 \label{eq:charge_retarded_phase}
\end{align}
Bob is switched on before Alice, with a propagation margin, and remains static until Alice has closed and her final field change has reached him. If Alice has also been static for a light-travel time before crossing,
\begin{align}
 \int_{-\infty}^0\mathcal K_\tau f_A(s)\,ds
 &=\int_{-\infty}^{-\tau}f_A(s)\,ds+\tau\notag\\
 &=\int_{-\infty}^0 f_A(s)\,ds.
 \label{eq:charge_retarded_identity}
\end{align}
Thus $\phi_{AB}=\phi_{BA}$ at crossing. The equality also holds after the complete pulse, when the derivative terms integrate to zero. For nonoverlapping spherical profiles, the harmonic mean-value property preserves the static kernel $k_e/d^3$; applying the endpoint identity to each source-point pair replaces the settling margin by $\tau_+=(d+a_A+a_B)/c$. Only post-crossing outward influence vanishes; the pre-crossing contribution to $\phi_{AB}$ remains.

Initialize $|++\rangle$ with zero controlled dipoles and the local electromagnetic vacuum. At the reciprocal-phase endpoints, the Gaussian field trace gives \cite{ChargeSugiyama2023}
\begin{align}
 \rho_{s,s'}&=\frac14 e^{-\Gamma_A d_A^2-\Gamma_B d_B^2-\Gamma_c d_A d_B}
 e^{-i\Phi(s_As_B-s'_As'_B)/2},\notag\\
 d_j&=(s_j-s'_j)/2,\qquad s_j,s'_j=\pm1,\notag\\
 \Gamma_j&\ge0,\qquad |\Gamma_c|\le2\sqrt{\Gamma_A\Gamma_B}.
 \label{eq:charge_gaussian_state}
\end{align}
Here $\Gamma_j=\langle\varphi_j^2\rangle/2$, $\Gamma_c=\langle\{\varphi_A,\varphi_B\}\rangle/2$, with $\varphi_j=(2p_j/\hbar)\int f_jE_{x,u_j}\,dt$. The real exponential is a joint Gaussian characteristic function, hence a mixture of local $Z$ rotations. A cubic turn-on of duration $T_{\rm on,A}$ followed by a stationary hold $H>\tau_+$ gives
\begin{equation}
 \Phi(0)=\frac{2k_ep_Ap_B}{\hbar d^3}
       \left(\frac{T_{\rm on,A}}2+H\right)=m\pi.
 \label{eq:charge_crossing_separable}
\end{equation}
Choose an integer $m$ satisfying this condition. Since $e^{-im\pi Z_AZ_B/2}$ is local up to a global phase, the crossing state is separable even after tracing the field. Its static dressing correlations need not be small; crossing separability does not imply field independence.

\textit{Continued Coupling and Radiation.}---After crossing, Alice holds for $t_h=x\tau$ and closes over $T_A=y\tau$, using $f_A=1-3u^2+2u^3$, $u=(t-t_h)/T_A$. Bob remains coupled and closes slowly only after $t_h+T_A+\tau_+$. Up to the calibrated local crossing phase, the final parity phase is
\begin{equation}
 \Phi_{\rm em}=\frac{2k_ep_Ap_B}{\hbar d^3}
                       \left(t_h+\frac{T_A}{2}\right).
 \label{eq:charge_continued_phase}
\end{equation}
Alice samples only Bob's pre-crossing field, while Bob subsequently receives her final field change. No post-crossing influence propagates outward from Bob.

For a cubic switching edge of duration $T_e$, the overlap exponent of the emitted coherent radiation states is half the photon number of the difference source $\Delta p=2pf$:
\begin{equation}
 \Gamma_{\rm edge}=\frac{k_e}{3\pi\hbar c^3}
  \int_0^\infty\frac{|\Delta\widetilde{\ddot p}(\omega)|^2}{\omega}\,d\omega
  =\frac{12k_ep^2}{\pi\hbar c^3T_e^2}.
 \label{eq:charge_radiation_edge}
\end{equation}
The dimensionless integral is $\int_0^\infty k|\widetilde w(k)|^2dk=9$ for $w(u)=6u(1-u)$ on $(0,1)$. Finite smearing can only reduce this norm because $|\widetilde u_j|\le1$. Opening and closing radiation add as amplitudes, so the triangle inequality, without assuming independent stages, gives
\begin{equation}
 \Gamma_j\le
 \left(\sqrt{\Gamma_{j,\rm on}}+\sqrt{\Gamma_{j,\rm off}}\right)^2.
 \label{eq:charge_radiation_bound}
\end{equation}
These are final closed-source exponents, not the intermediate static dressing in Eq.~\eqref{eq:charge_gaussian_state}.

Put $g_j=(2p_j/d)\sqrt{k_e/(\hbar c)}$, $T_{\rm on,A}=P\tau$ and $T_{\rm on,B}=T_{\rm off,B}=Q\tau$. Then
\begin{align}
 \Phi_{\rm em}&=\frac{g_Ag_B}{2}(x+y/2),\notag\\
 \Gamma_A&\le\frac3\pi g_A^2(P^{-1}+y^{-1})^2,
 &\Gamma_B&\le\frac{12g_B^2}{\pi Q^2}.
 \label{eq:charge_noise_budget}
\end{align}
Increasing Bob's dipole strengthens the phase without increasing Alice's radiation; his longer switching time compensates his larger radiation cost.

\textit{Finite Benchmark and One-Way Distillation.}---Choose
\begin{gather}
 g_A=0.1,\quad g_B=20\pi,\quad P=100,\quad Q=10^4,\notag\\
 x=0.2,\quad y=0.6,\quad H=2\tau.
 \label{eq:charge_benchmark}
\end{gather}
Alice opens on $[-102,-2]\tau$ and closes on $[0.2,0.8]\tau$. Bob opens on $[-10104,-104]\tau$ and closes on $[2,10002]\tau$. Taking $a_A+a_B\ll0.2d$ leaves propagation and readout margins. The resulting crossing phase is $52\pi$, with
\begin{equation}
 \Phi_{\rm em}=\pi/2,\qquad
 \Gamma_A\le0.0268451,\quad\Gamma_B\le0.000150797.
 \label{eq:charge_benchmark_output}
\end{equation}
The complete history requires $R_s\gg2.01\times10^4d$, not merely $R_s\gg d$. These are controlled, potentially mesoscopic polarization parameters; gravitational branch effects, mutual-force recoil and controller records must remain within the error budget.

At $\Phi_{\rm em}=\pi/2$, Bob's marginal is maximally mixed. For arbitrary allowed $\Gamma_c$, the fidelity with the ideal parity state obeys
\begin{align}
 F&=\frac{1+e^{-\Gamma_A}+e^{-\Gamma_B}
       +e^{-\Gamma_A-\Gamma_B}\cosh\Gamma_c}{4}\notag\\
 &\ge1-\epsilon,\qquad \epsilon=(\Gamma_A+\Gamma_B)/2.
\end{align}
Since $S(\rho_{AB})\le h_2(\epsilon)+\epsilon\log_2 3$ for $\epsilon\le3/4$, the one-way hashing inequality \cite{Devetak2005} gives
\begin{align}
 E_{\rightarrow}&\ge I(A\rangle B)=S(\rho_B)-S(\rho_{AB})\notag\\
 &\ge1-h_2(\epsilon)-\epsilon\log_2 3
 \ge0.8754\ \text{ebits/use}.
 \label{eq:charge_oneway_rate}
\end{align}
This is an asymptotic rate for independent copies in the specified model, before Bell-analyzer losses, not a single-trial success probability. It retains cross-party radiation covariance and requires no photon collection or parity measurement.

\textit{Closure and Exterior Transfer.}---Coherent confinement must close the full charge and motional configurations, not merely cancel their leading dipoles. For a local harmonic coordinate $X$ with conditional force $sF(t)$, choosing $F=M(\ddot X_c+\Omega^2X_c)$ makes its mean follow $sX_c$; $X_c=\dot X_c=0$ at both endpoints closes its conditional displacement. Residual controller and mutual-force records enter the output state. The benchmark has a finite margin: an additional trace-distance error $\delta=0.01$ reduces its coherent-information lower bound by at most $2h_2(\delta)+\delta\log_2 3$, leaving $I>0.69$. Smooth endpoint regularization is likewise allowed within this margin.

For point centers the horizon is $z=ct$, with $z_B=0$ and $z_A=d$. For finite sources, define $t=0$ as completion of Bob's crossing and shift it to $z=ct+a_B$; Alice remains wholly outside until $(d-a_A-a_B)/c$. After closing at $0.8\tau$, she transfers her internal qubit to an exterior register using an auxiliary EPR pair and a local Bell measurement \cite{Bennett1993}. Emission of the Bell record before this deadline lets a register at $z=2d$ receive it before its own crossing and launch the output carrier outward.

After Alice decouples, her local operations commute with all subsequent Bob--field evolution. Her early transfer therefore gives the same final exterior--Bob state as transfer after Bob's slow closure; an isolated entangled pair need not already exist at her Bell measurement. For a preselected finite distillation block, Alice can likewise perform the sender operations before her deadline, send the syndrome inward, and transfer the retained registers outward. Bob stores the syndrome and completes his operations after closure. Choose laboratory and black-hole scales to accommodate the finite block and processing; no reply from Bob is required. The Bell-analyzer yield is implementation-dependent.

This construction establishes continued electromagnetic phase interaction and eventual distillable entanglement within the finite-source, locally flat QED model. It does not by itself verify the gravitational field model. The horizon merely constrains chronology and imposes one-way classical communication.

\section{Appendix C: Phase During Optical Recombination}
\label{app:recombination}
Set $d=d_{\rm eq}$, $b=\Delta x/2$, and $\alpha=Gm_{\rm br}^2/(\hbar c)$. For $0\le u\le b/c$, Alice's packets follow $x_A=\pm(b-cu)$ at $z=d$, with Bob at $(\pm b,0)$, giving
\begin{equation}
R_\parallel^2=d^2+c^2u^2,\qquad
R_\times^2=d^2+(2b-cu)^2.
\end{equation}
For Bob's prescribed static monopole $U=-Gm_{\rm br}/R$, the weak-field metric $g_{00}=-(1+2U/c^2)$, $g_{ij}=(1-2U/c^2)\delta_{ij}$ gives $H_\gamma=cp(1+2U/c^2)$ to first order. With $cp=m_{\rm br}c^2$, the fixed-reference propagation phase is
\begin{align}
\Phi_\gamma
&=\frac{2Gm_{\rm br}^2}{\hbar}\int_0^{b/c}
\left(\frac1{R_\parallel}-\frac1{R_\times}\right)du\notag\\
&=2\alpha\left[2\operatorname{arsinh}(r/2)-\operatorname{arsinh}r\right].
\label{eq:optical_recombination_phase}
\end{align}
The scalar Newtonian pair potential gives half this contribution. At $r=r_\star$, $m_{\rm br}=m_p/2$, $\Phi_\gamma=0.047239901\ldots$~rad, or $0.084113872\ldots$~rad with the hold phase, before controller and endpoint contributions. Since $R\ge d$ and $ct<d$, all sampled source times have $t_{\rm ret}\le0$. Launch, recoil, and mode conversion remain outside this propagation calculation. Reflection-even logical profiles at the symmetry plane couple identically to Bob's branches during projected readout, adding no joint phase; earlier controller records still require coherent readout.

\end{document}